\documentclass{iopjournal}
\usepackage{xspace}
\usepackage{graphicx}
\usepackage[T1]{fontenc}
\usepackage{lmodern}
\usepackage{amsmath} 
\usepackage{amssymb} 
\usepackage[numbers,sort&compress]{natbib}
\usepackage{fontawesome}

\newcommand{\Msol}{\ensuremath{\rm M_{\rm \odot}}\xspace}
\newcommand{\de}{{\rm d}}

\newcommand{\gsirens}{\texttt{Gsirens}\xspace}

\newcommand{\hu}{\ensuremath{{\rm km s^{-1} Mpc^{-1}}}\xspace}

\begin{document}

\articletype{Article} 

\title{Non-Parametric Reconstruction of the Hubble Parameter from the Fourth Gravitational Wave Transient Catalog and DESI Baryonic Acoustic Oscillations}

\author{Grégoire Pierra$^{1,*}$\orcid{0000-0003-3970-7970}, Alberto Colombo$^{1,2}$\orcid{0000-0002-7439-4773} and Simone Mastrogiovanni$^1$\orcid{0000-0003-1606-4183}}

\affil{$^1$INFN, Sezione di Roma, 1-00185 Roma, Italy}

\affil{$^2$INAF – Osservatorio Astronomico di Brera, via Emilio Bianchi 46, I-23807 Merate (LC), Italy}

\affil{$^*$Author to whom any correspondence should be addressed.}

\email{gregoire.pierra@roma1.infn.it}


\begin{abstract}
The release of the fourth Gravitational-Wave Transient Catalog (GWTC-4.0) by the LIGO–Virgo–KAGRA collaboration includes more than 200 compact binary coalescence (CBC) candidates that can be used to probe the cosmic expansion. The population of merging binary black holes has been used so far to provide a constraint on the Hubble constant and dark matter fraction under the hypothesis of a flat-$\Lambda$-Cold-Dark-Matter Universe.

In this work, we provide the first non-parametric constrain on the Hubble parameter from 137 dark sirens reported in GWTC-4.0.
We employ the relation between detector and source frame masses for detected GW signals, to obtain a statistical redshift evaluation for the population of binary black holes (BBHs). We model the Hubble parameter as a non-parametric autoregressive process in terms of the scale factor, using splines. In addition, we introduce two novel features: the use of \textit{anchor} points for $H(z)$ derived from an external probe — here, Baryon Acoustic Oscillations (BAOs) — and a constraining power coefficient that quantifies where the inference is most data-driven by GW detections.

We highlight three key findings: (i) using GWs alone, the Hubble parameter determination is the most GW-data-driven around redshift $z = 0.44$, yielding to $H(0.44) = 92.3_{-36.6}^{+29.9}\,\hu$. Its value at $z = 0$, the Hubble constant, is therefore less constrained by the GW data. (ii) The Hubble parameter inferred from analyses assuming a flat-$\Lambda$CDM cosmological model is strongly affected by the cosmological model assumption. (iii) Introducing an anchor point for $H(z)$ enhances the inferred constraints and provides a clear visualization of the redshift range where GWs contribute most to the constraining power.
\end{abstract}

\section{Introduction \label{section: introduction}}
Measuring the cosmic expansion history of the Universe is a central goal of modern cosmology, with a particular focus on determining its local expansion rate ($H_0$), the Hubble constant. This question has attracted considerable attention due to the growing discrepancy between early--universe estimates of $H_0$ inferred from the cosmic microwave background (CMB) \cite{Planck:2018vyg} and late--universe measurements based on standardized candles such as type~Ia supernovae \cite{Riess:2021jrx}. This discrepancy, widely referred to as the ``Hubble tension'', has now surpassed the $5\sigma$ level.

Gravitational waves (GWs) from compact binary coalescences (CBCs) have recently emerged as an independent and powerful probe of the Hubble constant, complementing both CMB and standard-candle-based methods. The concept of using GWs as standard sirens, by analogy with standard candles in cosmology, was first proposed by Schutz in 1986 \cite{Schutz:1986gp} and has since been extended in numerous studies when an electromagnetic counterpart is present \cite{Markovic:1993cr, Dalal:2006qt, Feeney:2018mkj, Mancarella:2024qle}, and when not \cite{Schutz:1986gp, Holz:2005df, Taylor:2012db, You:2020wju, Mancarella:2021ecn, Ye:2021klk,  Ezquiaga:2022zkx, Mastrogiovanni:2023emh, Gray:2023wgj, Oguri:2016dgk, Mukherjee:2020hyn, Bera:2020, Dalang:2024gfk,Mali:2024wpq, MaganaHernandez:2025cnu}. GW observations provide a direct measurement of the luminosity distance to the source, independent of the traditional distance ladder, and when paired with a redshift measurement, enable constraints on cosmological parameters such as the Hubble constant. In this paper, we focus on the ``spectral siren'' method \cite{Mastrogiovanni:2021wsd,Ezquiaga:2022zkx}, a technique able to exploit the difference between detector and source mass and obtain a redshift estimation for the population of GW sources.

While determining $H_0$ is important, an even more compelling objective is to chart the expansion history of the Universe through the Hubble parameter, $H(z)$. Measuring $H(z)$ as a function of redshift allows us to constrain cosmological models, investigate the evolution of the expansion, and potentially probe the nature of dark energy and dark matter. GWs from binary black hole observations are excellent sources for this task as they are observed up to redshift 2 with current detectors sensitivities.
Current cosmological expansion measurements from GW sources typically adopt flat-$\Lambda$CDM framework \cite{DES:2020nay, Palmese:2020, LIGOScientific:2021aug, Mukherjee:2022afz, Bom:2024afj, MaganaHernandez:2025cnu, LIGOScientific:2025jau}, and thus the reconstructed behavior of $H(z)$ is implied by the model assumption and only constraints on $H_0$ or $\Omega_m$ can be reported.
With the recent release of the largest GW catalog to date by the LIGO–Virgo–KAGRA (LVK) collaboration \cite{VIRGO:2014yos, LIGO:2024kkz, LIGOScientific:2025hdt}, the Gravitational-Wave Transient Catalog~4.0 (GWTC-4.0) \cite{LIGOScientific:2025yae,LIGOScientific:2025slb, LIGOScientific:2025hdt}, which contains a total of 218 candidate detections from CBCs, the potential of standard sirens for cosmological studies can now be explored in unprecedented detail. 

We present the first non-parametric spectral siren inference of the Hubble constant, and more generally of the Hubble parameter $H(z)$, using GW observations. Specifically, we analyze 137 BBHs with a false-alarm rate (FAR) below $0.25\,\mathrm{yr}^{-1}$, as reported in GWTC-4.0 \cite{Abac:2025saz}, modeling the Hubble parameter as an autoregressive process with splines. We further assess the extent to which GW data themselves constrain the expansion history across redshift, compared with the constraints induced by model assumptions in parametric approaches. 
Our proposed model further allows for an agnostic combination of GW sources detected across a broad luminosity-distance range with independent constraints on the Hubble parameter $H(z)$ from other cosmological probes. In this work, we specifically investigate the effect of incorporating external $H(z)$ measurements, for instance those provided by Baryonic Acoustic Oscillations (BAOs) from DESI \cite{DESI:2024mwx}.
Some studies have also investigated the use of BAO measurements in the context of testing modified gravity theories, although used in a different framework with a different application from the one explored in this work \cite{Mukherjee:2020mha, Afroz:2023ndy, Afroz:2024joi}.

The paper is organized as follows. In Section~\ref{section: method}, we present our new approach for constructing a non-parametric model of the Hubble parameter using splines and discuss its connection to the equation of state of the Universe. Section~\ref{section: framework} introduces the Bayesian framework used to perform the spectral siren inference, along with the population models and pipeline employed for the analyses. Section~\ref{section: results} summarizes our main findings, including the non-parametric estimation of the Hubble parameter from GWs, with and without anchors from BAOs, and comparison with the widely used flat-$\Lambda$CDM cosmological model. It also quantifies the effect of using parameterized cosmological models for spectral sirens and identifies when the inference of $H(z)$ is genuinely driven by the GW data. Finally, we conclude in Section~\ref{section: conclusion}.

\section{A non-parametric cosmological model \label{section: method}}
The spectral siren method estimates the cosmological redshifts of GW sources by exploiting the intrinsic relation between the detector mass $m_{\rm d}$ and the source mass $m_{\rm s}$:
\begin{equation}
    m_{\rm d} = (1+z) \, m_{\rm s},
    \label{eq:mass_redshift_degeneracy}
\end{equation}
where $z$ is the \emph{cosmological redshift} of the source, and $m_{\rm d}$ is commonly referred to as the \emph{redshifted mass}. In practice, however, the redshift of a GW source is not directly measured from the signal, unlike the luminosity distance, but is instead inferred statistically. This is achieved by fitting the distribution of BBHs in the luminosity distance–detector mass space, assuming a parametric source mass model and using the cosmological evolution encoded in the Friedmann equations.

\begin{figure}[ht]
 \centering
        \includegraphics[width=0.95\textwidth]{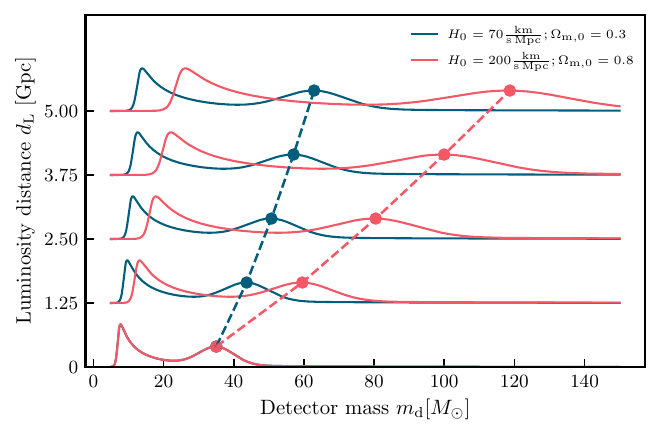}
 \caption{Graphical representation of the evolution of the mass spectrum in the detector frame as observed at increasing luminosity distances, assuming different values of the Hubble constant ($H_0$) and the matter-density parameter ($\Omega_{\rm m,0}$) within a flat-$\Lambda$CDM cosmological model. The figure illustrates how changes in cosmological parameters affect the observed shape of mass features at larger distances.}
\label{fig:mass_spectrum_evolution}
\end{figure}

Figure~\ref{fig:mass_spectrum_evolution} illustrates the procedure underlying the spectral siren inference method. It shows the evolution of two mass spectra in the luminosity distance–detector-frame mass plane, assuming a flat-$\Lambda$CDM cosmological model with two different choices for the Hubble constant and matter-density parameters. For nearby sources ($d_{\rm L} \approx 0~\mathrm{Mpc}$), i.e., GW signals emitted close to the observer, the detector mass spectra coincide for both cosmologies. This is expected because, according to Eq.~\ref{eq:mass_redshift_degeneracy}, the source mass equals the detector mass at vanishing redshift. As sources are pushed to larger distances, the detector mass spectrum evolves, reflecting its dependence on the cosmological redshift $z$. This evolution occurs because the source mass distribution is assumed to be redshift-independent, while the mapping between detector and source masses depends on the cosmological parameters. For illustration, the shift of the Gaussian peak initially located at $35~M_\odot$ for $d_{\rm L}=0~\mathrm{Mpc}$ highlights how the evolution of the mass spectrum varies with the adopted cosmological model and parameter values.

The evolution of the mass spectrum for a given cosmology is governed by the Friedmann equations, and in particular by the redshift dependence of the Hubble parameter $H(z)$, since
\begin{equation}
    m_{\rm d} = (1+z)\,m_{\rm s}
    = \left[1 + f(d_{\rm L}; H(z))\right] m_{\rm s},
    \label{eq:mass_redshift_degeneracy_parametrized}
\end{equation}
where $f(\cdot)$ denotes the mapping between luminosity distance and cosmological redshift for a specified cosmological model, here assumed to be a flat-$\Lambda$CDM Universe. Throughout the remainder of this study, we adopt this flat-$\Lambda$CDM framework as the fiducial model, or stated otherwise when the non-parametric approach is employed.
In the low-redshift regime, the function $f(\cdot)$ can be approximated as
\begin{equation}
    f(d_{\rm L}; H(z{=}0)) \approx \frac{d_{\rm L} H_0}{c},
    \label{eq:f_low_z_limit}
\end{equation}
where $c$ is the speed of light and $H_0$ denotes the present-day value of the Hubble parameter, the Hubble constant. A direct consequence within flat-$\Lambda$CDM cosmological models is that higher values of $H_0$ lead to a more pronounced evolution of the detector mass spectrum, as illustrated in Fig.~\ref{fig:mass_spectrum_evolution}. In the low-redshift limit, however, the matter-density parameter $\Omega_{\rm m,0}$ plays no significant role, as indicated by Eq.~\ref{eq:f_low_z_limit}.  
In spectral siren cosmology, the constraining power of this approach arises from the fact that, for a given luminosity distance, different cosmological models predict different redshifts through Eq.~\ref{eq:mass_redshift_degeneracy_parametrized}, resulting in distinct detector-frame mass distributions and evolutionary trends.

In this study, we aim to develop a non-parametric method to infer the cosmic expansion history solely from GW observations. Accordingly, it is natural to represent the cosmological model in a non-parametric form by mapping selected redshift values to corresponding luminosity distance nodes or bins.  
Since GW signals are observed in the luminosity distance–detector mass parameter space, this framework allows us to identify which GW candidates contribute to the estimation of the Hubble parameter, now expressed as a function of luminosity distance, $H(d_{\rm L})$. These non-parametric measurements can then be compared \emph{a posteriori} with the predictions of standard cosmological models, such as the flat-$\Lambda$CDM Universe.

However, adopting a non-parametric model of luminosity distance as a function of redshift introduces significant challenges. First, the spectral siren method requires the Hubble parameter to estimate the rate density of CBC mergers in the Universe. A non-parametric representation of $z(d_{\rm L})$ can, in some cases, yield negative values of the Hubble parameter, even if $z(d_{\rm L})$ is constrained to be monotonic. Such behavior would correspond to a Universe that contracts rather than expands, which is physically inconsistent with current observations. Moreover, certain non-parametric functions, such as splines \cite{Callister:2023tgi, Edelman:2021zkw, Edelman:2022ydv, Golomb:2022bon}, one of the simplest and most commonly used non-parametric representations, can produce discontinuities in $H(z)$ at the redshift nodes.  
Second, depending on how the function $z(d_{\rm L})$ is constructed, the method may explore cosmologies with extreme or otherwise unphysical values of $H(z)$, for instance in cases where $z(d_{\rm L})$ evolves very slowly with distance.

Given the challenges discussed above, we construct our non-parametric model directly for the Hubble parameter, $H(z)$, as a function of redshift, using a spline parameterization in natural logarithmic space. The cosmological redshift range is divided into $N_{\rm nodes}$ redshift nodes, $z_i$, and the natural logarithm of the Hubble parameter is then defined such that
\begin{equation}
    \ln H(z) = \sum_{i=1}^{N_{\rm nodes}-1} \left[ a_i \ln\left(\frac{1+z}{1+z_i} \right) +b_i \right]\Theta(z;z_{i},z_{i+1}),
    \label{eq:Hz_sp}
\end{equation}
where $\Theta(\cdot)$ is a box function ensuring no contribution to the Hubble parameter from intervals outside $[z_i, z_{i+1}]$. 
From Eq.~\ref{eq:Hz_sp}, the coefficients $a_i$ and $b_i$ can be expressed as
\begin{eqnarray}
    a_i&=& \ln\left[\frac{H(z_{i+1})}{H(z_{i})} \right]/ \ln\left[\frac{1+z_{i+1}}{1+z_{i}} \right] \\
    b_i&=& \ln H(z_i).
\end{eqnarray}

This spline-based parameterization offers several advantages. First, the Hubble parameter constructed in this way is a continuous function, in contrast to the discontinuities encountered in other non-parametric approaches. Second, and more importantly from a physical perspective, the Hubble parameter now scales locally as $H(z) \propto (1+z)^{\alpha} = a^\alpha$, where $a$ is the scale factor thus providing a direct connection to the equation of state (EOS) of the Universe and its energetic content.
Indeed, under the hypothesis that a single energy component dominates within a redshift bin, the Hubble parameter can be rewritten as
\begin{equation}
    H(z) = H(z_i) \left[\frac{1+z}{1+z_i}\right]^{3(1+w_i)} = H(a_i) \left[\frac{a_i}{a}\right]^{3(1+w_i)},
    \label{eq:Hz_sp_eos}
\end{equation}
where, for example, $w_i=0$ corresponds to a Universe dominated by cold dark matter in the redshift bin $z_i$. 

Using the parametrization introduced in Eq.~\ref{eq:Hz_sp}, the comoving volume element between redshifts $z_i$ and $z_{i+1}$, for $z_i < z < z_{i+1}$, can be written as
\begin{equation}
\Delta d_{\rm com,i}(z) =
\begin{cases}
\frac{c (1+z_i)}{H(z_i)(1-a_i)}(1+z)^{1-a_i} \left[\frac{1+z_{i+1}}{1+z} \frac{H(z)}{H(z_{i+1})}-1\right] & \text{if } a_i \neq 1, \\
\frac{1+z_i}{H(z_i)} \ln \left[\frac{1+z_{i+1}}{1+z} \right]  & \text{if } a_i = 1\,.
\end{cases}    
\end{equation}
This expression provides a convenient way to approximate the comoving distance by discretizing the redshift range into intervals defined by the nodes $z_i$. Within each interval, the Hubble parameter is approximated according to the chosen parametrization, allowing for an analytic expression for $\Delta d_{\rm com,i}(z)$ that accounts for the variation of $H(z)$ across the bin.
The total comoving distance to a redshift $z$ can then be obtained by summing the contributions from all nodes below $z$ and including the partial interval up to $z$ itself:
\begin{equation}
    d_{\rm com}(z) =\sum_{j=1}^{N_j-1} \Delta d_{\rm com, j}(z_{j+1}),
    \label{eq:comoving}
\end{equation}
where $N_j$ is the number of redshift nodes satisfying $z_j < z$. This approach effectively constructs the comoving distance by combining exact contributions within each redshift bin, which is particularly convenient for spectral siren analyses where the source redshifts are sampled discretely.
Therefore, Eqs.~\ref{eq:Hz_sp} and \ref{eq:comoving} provide all the cosmological ingredients needed to compute the comoving distances required in a spectral siren framework.

\section{The spectral siren inference framework \label{section: framework}}

\subsection{Hierarchical likelihood and constraint estimators}

To jointly infer the population and cosmological parameters of our GW catalog, we employ a hierarchical Bayesian inference (HBI) framework \cite{Mandel:2018mve, Vitale:2020aaz}. In particular, we focus on the spectral siren approach, which relies solely on GW data, without the use of external galaxy catalogs to estimate source redshifts. The detection of GW signals by terrestrial interferometers can be modeled as an inhomogeneous Poisson process, subject to selection effects arising from the finite sensitivity of the GW detectors.

Following \cite{Mandel:2018mve, Vitale:2020aaz}, the hierarchical likelihood of observing $N_{\rm obs}$ GW events described by a dataset $\{\mathbf{x}\}$ over an observation time $T_{\rm obs}$ can be written as:
\begin{eqnarray}
    \mathcal{L}(\{\mathbf{x}\}|\mathbf{\Lambda},H(z)) \propto e^{-N_{\rm exp}(\mathbf{\Lambda})}\prod^{N_{\rm obs}}_{i=1} T_{\rm obs} \int \de \boldsymbol{\theta}\de z\mathcal{L}(\mathbf{x}_{i}|\boldsymbol{\theta},z,\mathbf{\Lambda})\frac{1}{1+z}\frac{\de N_{\rm cbc}(\mathbf{\Lambda})}{\de \boldsymbol{\theta}\de z \de t_{\rm s}},
    \label{eq:likelihood}
\end{eqnarray}
where $\mathcal{L}(\mathbf{x}_{i}|\boldsymbol{\theta},z,\boldsymbol{\Lambda})$ is the likelihood of a single event, with $\boldsymbol{\theta}$ denoting the intrinsic parameters of the binary, $z$ the cosmological redshift, and $\boldsymbol{\Lambda}$ the population and cosmological hyper-parameters. The correction for selection effects is encoded in the expected number of detections, $N_{\rm exp}(\boldsymbol{\Lambda})$, while the final term in Eq.~\ref{eq:likelihood} represents the CBC merger rate.  

This hierarchical likelihood can also be expressed in an equivalent ``\textit{scale-free}'' form, given by:
\begin{eqnarray}
    \mathcal{L}(\{\mathbf{x}\}|\mathbf{\Lambda},H(z)) \propto \prod^{N_{\rm obs}}_{i=1} \frac{\int \de \boldsymbol{\theta} \de z \mathcal{L}(\mathbf{x}_{i}|\boldsymbol{\theta},z,\boldsymbol{\Lambda})\frac{1}{1+z}\frac{\de N_{\rm cbc}(\mathbf{\Lambda})}{\de \boldsymbol{\theta}\de z \de t_{\rm s}}}{\int \de \boldsymbol{\theta}\de z p_{\rm det}(\boldsymbol{\theta},z,\boldsymbol{\Lambda},H(z))\frac{1}{1+z}\frac{\de N_{\rm cbc}(\mathbf{\Lambda})}{\de \boldsymbol{\theta}\de z \de t_{\rm s}}},
    \label{eq:likelihood_scalefree}
\end{eqnarray}
where the probability of detection, $p_{\rm det}(\boldsymbol{\theta},z,\boldsymbol{\Lambda})$, accounts for selection effects. For further details on the derivation of this likelihood and the definitions of the various terms, we refer the reader to \cite{Mastrogiovanni:2023emh}.

In our analysis, the CBC population is described by the set of intrinsic parameters $\boldsymbol{\theta} = (m_{1,\rm s}, q)$, where the mass ratio is defined as $q = m_{2}/m_{1}$. Spins are not considered in this work. With this choice of population parameters, the hierarchical likelihood in Eq.~\ref{eq:likelihood_scalefree} can be rewritten by expressing the CBC merger rate as:
\begin{eqnarray}
    \mathcal{L}(\{\mathbf{x}\}|\mathbf{\Lambda},H(z)) &\propto& \prod^{N_{\rm obs}}_{i=1} \frac{\int \mathcal{L}(\mathbf{x}_{i}|m_{1,\rm s},q,z,\mathbf{\Lambda})\mathcal{R}(z;\boldsymbol{\Lambda})p_{\rm pop}(m_{1,\rm s},q|\boldsymbol{\Lambda})\frac{\de V_{\rm c}}{\de z}\frac{1}{1+z}\de m_{1,\rm s} \de q \de z}{\int p_{\rm det}(m_{1,\rm s},q,z,\boldsymbol{\Lambda},H(z))\mathcal{R}(z;\boldsymbol{\Lambda})p_{\rm pop}(m_{1,\rm s},q|\boldsymbol{\Lambda})\frac{\de  V_{\rm c}}{\de z}\frac{1}{1+z}\de m_{1,\rm s} \de q \de z } \nonumber \\
    &\propto& \prod^{N_{\rm obs}}_{i=1} \mathcal{Z}_i, 
    \label{eq:likelihood_scalefree_long}
\end{eqnarray}
where $\mathcal{R}(z;\boldsymbol{\Lambda})$ denotes the CBC merger rate density as a function of redshift, and $\frac{\de V_{\rm c}}{\de z}$ is the differential comoving volume. In general, the merger rate density can be expressed as the product of the local CBC merger rate, $\mathcal{R}_0$, and a redshift-dependent rate function, $\psi(z;\boldsymbol{\Lambda})$, which is typically modeled to follow the star formation rate \cite{Madau:2014bja}.
Importantly, Eq.~\ref{eq:likelihood_scalefree_long} highlights that the hierarchical likelihood is constructed as the product of the marginal likelihoods of individual events, $\mathcal{Z}_{i}$. From a cosmological perspective, this structure is central: cosmological information enters through the mapping between observed quantities and source-frame properties, encoded in $H(z)$ and the comoving volume element. This allows the spectral siren approach to extract cosmological constraints directly from the observed GW population, without relying on external redshift information.

The integrals appearing in the hierarchical likelihood of Eq.~\ref{eq:likelihood_scalefree_long} are evaluated via Monte Carlo integration \cite{MOROKOFF1995218, Talbot:2023pex}. This is done using a set of simulated detected GW signals (injections) to estimate the selection function statistically, together with posterior samples from the observed GW events, enabling an efficient numerical approximation of the integrals. To ensure numerical stability and avoid bias in these integrals, we apply two cuts on the effective number of samples: one on the posterior samples and another on the injections, set respectively to 10 and $4\,N_{\rm obs}$, following the same setup of \cite{LIGOScientific:2025jau}.

Inspired by the method presented in \cite{Mancarella:2024qle}, which proposes a way to quantify the contribution of individual GW events to the constraints on the Hubble constant or other population parameters, we construct a similar quantity in this study. Specifically, we define a parameter $\mathcal{C}_{k}$ representing the fraction of GW events contributing to the constraints on the Hubble parameter at a given redshift node $H(z_k)$. Accordingly, we define $\mathcal{C}_{k}$ as:
\begin{eqnarray}
    \mathcal{C}_k =\frac{1}{N_{\rm obs}} \frac{\sum^{N_{\rm obs}}_{i=1} |\rho(\mathcal{Z}_i,H(z_k))| \, {\rm Var}[\mathcal{Z}_i]}{\sum_{i=1}^{N_{\rm obs}} {\rm Var}[\mathcal{Z}_i]},
    \label{eq:coef_constraints}
\end{eqnarray}
where $\mathcal{Z}_i$ denotes the marginal likelihood of the $i$-th event from Eq.~\ref{eq:likelihood_scalefree_long}, and ${\rm Var}[\mathcal{Z}_i]$ is its variance. The term $\rho(\mathcal{Z}_i,H(z_k))$ is the Pearson correlation coefficient between that marginal likelihood and the Hubble parameter at redshift node $z_k$ \cite{72205df1-8a29-37a7-87c0-b4ebe1be7250}. This coefficient quantifies the extent to which a particular GW event $i$ contributes to the inference of a given parameter, here $H(z_k)$.
If this coefficient vanishes for a given event, it implies that variations in the population parameter have no effect on the single-event marginal likelihood. Consequently, the contribution of this event to the population likelihood is constant, and it does not influence the inference of that population parameter. Intuitively, GW events contribute to a given $H(z_k)$ if their marginal likelihood correlates with it; the stronger this correlation across the population, the tighter the constraints on the parameter\footnote{We note that $\mathcal{C}_k$ should not be interpreted strictly as the fraction of GW events contributing, as this would require a perfectly multivariate Gaussian posterior for $\mathcal{Z}_i$ and $H(z_k)$. However, it provides a useful indication of the most promising redshift ranges in terms of constraining power.
}. In Eq.~\ref{eq:coef_constraints}, both the Pearson correlation coefficient and the variance are calculated over the population posteriors. We note that Eq.~\ref{eq:coef_constraints} can be used for any population parameter $\Lambda$ on which we either set priors or it is implied by the population model, for instance the luminosity distance at a given redshift node.

\subsection{Population and cosmological models}
In addition to the non-parametric model with splines for the Hubble parameter, we must also define the population model for the mass distribution, $p_{\rm pop}(m_{1,\rm s},q|\boldsymbol{\Lambda})$, and the BBH merger rate, $\mathcal{R}(z;\boldsymbol{\Lambda})$, to construct the hierarchical likelihood in Eq.~\ref{eq:likelihood_scalefree_long}.

For the mass distribution, we adopt a model inspired by the fiducial BBH population presented in \cite{LIGOScientific:2025pvj}, namely a \textsc{broken power law + 2 peaks} model for the primary mass $m_{1,\rm s}$, coupled to a smoothed \textsc{power law} for the mass ratio $q$. The \textsc{broken power law + 2 peaks} model combines a broken power law $B$, with a slope transition at a break mass $m_{\rm break}$, with two truncated Gaussian peaks $G_1$ and $G_2$.
The full distribution for the primary mass, $\pi(m_{1,\rm s}|\boldsymbol{\Lambda})$, is then given by
\begin{eqnarray}
\label{eq: BPL 2p m1}
    p(m_{1,\rm s}|\boldsymbol{\Lambda}) = (1-\lambda_{\rm g})B(m_{1,\rm s}|\boldsymbol{\Lambda}) +\lambda_{\rm g}\lambda_{\rm g}^{\rm low} G_{1}(m_{1,\rm s}|\boldsymbol{\Lambda}) + \lambda_{\rm g}(1-\lambda_{\rm g}^{\rm low})G_{2}(m_{1,\rm s}|\boldsymbol{\Lambda})\,,
\end{eqnarray}
where $\lambda_{\rm g}$ and $\lambda_g^{\rm low}$ are the mixing fractions for the Gaussian peaks.
The mass distribution is defined between a minimum mass $m_{\rm min}$ and a maximum mass $m_{\rm max}$. Additionally, two smoothing factors $\delta_{\rm m,min}$ and $\delta_{\rm m,max}$ measured in solar masses and representing the width of a sigmoid function are applied to the lower and upper end of the spectrum to encode a possible tapering of the spectrum boundaries.
The mass ratio distribution is modeled as a power over the secondary mass and conditioned on the primary mass. Explicitly, the mass ratio distribution $\pi(q|\boldsymbol{\Lambda})$ reads
\begin{eqnarray}
    \pi(q|\boldsymbol{\Lambda}) \propto  q^{\beta_{q}} \Theta[m_{\rm min}/m_1,1],
\end{eqnarray}

For the BBH merger rate, $\mathcal{R}(z;\boldsymbol{\Lambda})$, describing the evolution of BBH mergers with cosmological redshift, we adopt the commonly used parameterization following the star formation rate from \cite{Madau:2014bja}:
\begin{eqnarray}
    \mathcal{R}(z;\boldsymbol{\Lambda}) = \big[1+(1+z_{\rm p})^{-\gamma - \kappa} \big]\frac{(1+z)^{\gamma}}{1+[(1+z)/(1+z_{\rm p})]^{\gamma +\kappa}},
\end{eqnarray}
where $z_{\rm p}$ represents the turning point between two regimes governed by the power indices $\gamma$ and $\kappa$.

Regarding the cosmological models, we consider two parameterizations: our main non-parametric model, where $H(z)$ is constructed using splines as an autoregressive function in log space (Section.~\ref{section: framework}), and the parametric flat-$\Lambda$CDM model, given by
\begin{eqnarray}
    H(z) = H_0\sqrt{\Omega_{\rm m,0}(1+z)^3 + (1-\Omega_{\rm m,0})},
\end{eqnarray}
which is commonly used in GW cosmology analyses.

For the autoregressive $H(z)$ model, we employ 100 redshift nodes, uniformly distributed in $\ln(1+z)$ up to $z_{\rm max}=10$. The spacing between consecutive nodes $i+1$ and $i$, denoted $\Delta u_i$, is
\begin{equation}
    \Delta u_i = \ln(1+z_{i+1}) - \ln(1+z_i) \approx 0.025,
\end{equation}
ensuring sufficient resolution in redshift. Between nodes, the Hubble parameter is interpolated using Eq.~\ref{eq:Hz_sp}. It follows that evolution of $H(z)$ between two nodes is
\begin{eqnarray}
    H(z_{i+1}) = H(z_i)\left(1 + \frac{3}{2}(1 + w_i \Delta u_i)\right),
    \label{eq:AR model}
\end{eqnarray}
where $w_i$ is the EOS parameter in the $i$-th redshift bin. We adopt a uniform prior $w_i \in [-2, 1/3]$, encompassing scenarios from phantom dark energy-dominated universes to radiation-like EOS.

In this non-parametric framework, the cosmological parameters consist of the Hubble parameter at $z=0$, $\ln H_0$, and the $(N_{\rm nodes}-1)$ EOS coefficients $w_i$. Table~\ref{tab:1} summarizes the priors and provides a description for each parameters (including the mass and rate prescriptions). Notably, in the limit $\Delta u_i \ll 1$, the autoregressive model of Eq.~\ref{eq:AR model} approximates the Hubble parameter in Eq.~\ref{eq:Hz_sp_eos}.
The flat-$\Lambda$CDM model is more straightforward in terms of parameterization and is described in detail in \cite{LIGOScientific:2025jau}.

\subsection{Data: GW catalog, injections, and BAO measurements}
All results presented in this study are based on the recent GWTC-4.0 catalog release by the LVK collaboration, which includes CBCs detected from the first observing run (O1) through the first part of the fourth observing run (O4a) \cite{LIGOScientific:2025slb}.
From the 218 GW candidates, we select a subset of 137 BBH events with a FAR $< 0.25\, \rm{yr}^{-1}$ to ensure that the majority of candidates are of astrophysical origin. We also exclude triggers occurring during the engineering run immediately preceding O4a, maintaining consistency with the latest LVK cosmological study \cite{LIGOScientific:2025jau}. The very massive event GW231123\_135430 \cite{LIGOScientific:2025rsn} is also excluded from our BBH sample, as key astrophysical properties, such as the component masses and inferred luminosity distance, appear to be highly sensitive to the waveform model choice during parameter estimation (PE).

We emphasize that the event selection in this study is based solely on the FAR threshold. In particular, we use the stable-release version of the GWTC-4.0 catalog, which is also employed in both population and cosmological analyses by the LVK \cite{LIGOScientific:2025jau,LIGOScientific:2025pvj}.
The GW detection sensitivity appearing in the denominator of Eq.~\ref{eq:likelihood_scalefree} is estimated using an injection campaign. Simulated GW signals (injections) are added to the real detector noise and recovered by the LVK search pipelines, and the evolving sensitivity for each observing run is accounted for directly in the injection file. For further details on the production and use of these injections in the Bayesian inference framework, we refer the reader to \cite{Essick:2025zed, Mastrogiovanni:2023zbw, Pierra:2025fgr}.

In addition to the GW-based approach, we complement our analysis by incorporating measurements of the Hubble parameter, $H(z)$, derived from BAOs. We use the latest BAO results from the Dark Energy Spectroscopic Instrument (DESI) \cite{DESI:2025zgx}, based on Data Release 2 (DR2) after three years of observations. BAOs provide a powerful tool for measuring the expansion history, using a characteristic scale imprinted on matter clustering by pressure waves that propagated in the coupled baryon-photon fluid of the pre-recombination Universe. 
The DESI DR2 sample includes over 14 million galaxy and quasar redshifts, spanning six tracer samples (e.g., luminous red galaxies, emission-line galaxies, quasars). These measurements provide distance constraints in the form of the transverse comoving distance, $D_\mathrm{M}(z)/r_\mathrm{d}$, normalized by the sound horizon at the drag epoch, $r_\mathrm{d}$, and the normalized radial distance, $D_\mathrm{H}(z)/r_\mathrm{d} \equiv (c/H(z))/r_\mathrm{d}$. For the calculation of $H(z)$, we adopt the value of the sound horizon at the drag epoch as estimated in \cite{Planck:2018vyg}.

\section{Results \label{section: results}}
All the results presented in this article were obtained using \href{https://baltig.infn.it/gravitysirens/gsirens_public/}{\gsirens~\faGithub}, a new Python package developed for rapid Bayesian inference with GW data. It relies on Hamiltonian Monte Carlo, specifically the No-U-Turn Sampler (NUTS) implementation in NumPyro and JAX \cite{jax2018github, phan2019composable}. The inference, including the hierarchical Bayesian analysis, was performed on a single NVIDIA GeForce RTX 4050 GPU. With this setup, analyzing the 137 BBH events, using 10,000 samples per event, takes approximately 10 hours with our spline model featuring 100 redshift nodes, and only 2 hours with the commonly used flat-$\Lambda$CDM cosmological model. 

For all the results on the reconstructed mass and mass ratio distributions and BBH merger rate as function of redshifts we obtain consistent results with \cite{LIGOScientific:2025jau}. In the following, we will focus on the constraints on the Hubble parameter.

\subsection{The Hubble parameter with GWs \label{sec:GWs}}
\begin{figure}[ht!]
    \centering
     \includegraphics[width=1.0\linewidth]{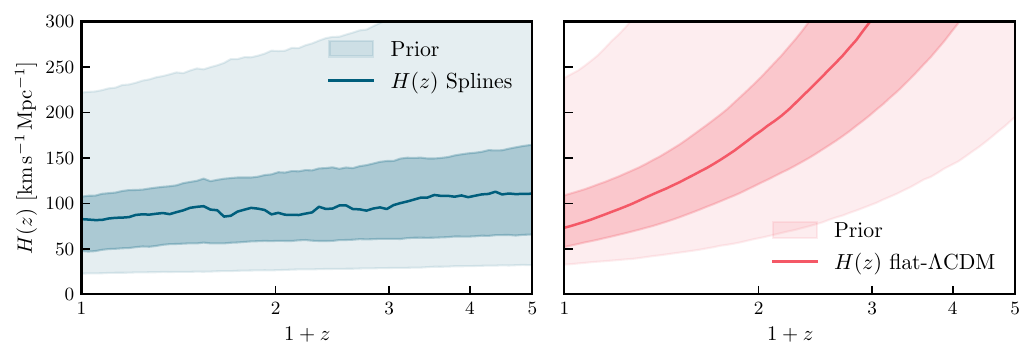}
    \caption{Prior and posterior predictive distributions of the Hubble parameter $H(z)$ as a function of cosmological redshift in $\rm log_{10}$ space, from a spectral siren inference using 137 GW detections with FAR$< 0.25\,yr^{-1}$, from the GWTC-4.0 catalog. \textbf{Left: }Non-parametric inference with the spline $H(z)$ model. Binned constraints are plot as continuous to improve readability. \textbf{Right: }Parametric inference of the Hubble parameter $H(z)$ assuming a flat-$\Lambda$CDM cosmological model. The dark colored contours represent the 68.3\% highest density intervals (HDI) around the maximum a posteriori (MAP) estimation.}
    \label{fig:Hz_GW_and_flat_LCDM}
\end{figure}

Figure.~\ref{fig:Hz_GW_and_flat_LCDM} presents the main result of this paper. The left panel shows the maximum a posteriori (MAP) estimation and 68.3\% highest density intervals (HDI) of $H(z)$ in each of the $i$-th redshift bins generated by the autoregressive model. The right panel displays the reconstructed projection of the Hubble parameter, this time assuming the standard parametric flat-$\Lambda$CDM cosmological model.

In both the parameterized and non-parametric reconstructions, it is clear that GW data encode some level of information about the Hubble parameter, as the inferred values are constrained well within their implied prior ranges. A perfectly non-informative inference would yield broader HDIs, covering the entire prior range, behavior which is not observed here. With both models, we obtain consistent estimates for the local value of the Hubble parameter, in agreement with current literature \cite{Planck:2018vyg, Riess:2021jrx, DESI:2024mwx, LIGOScientific:2025jau}. 
We find that the Hubble constant is measured at $H_0=85^{+38}_{-29}\,\hu$ and $H_0=86^{+42}_{-24}\,\hu$ for the non-parametric and parametric approaches, respectively.
While the reconstructions of the Hubble parameter are the same at low redshifts, they diverge significantly at higher redshifts. This suggests that the constraints on $H(z)$ and its behavior in this region, are driven primarily by the choice of the assumed cosmological model and its original priors, rather than being data-driven, as one might hope. This effect is further amplified by the fact that beyond a certain redshift ($z\sim1.2$), no GW detections are available, making the shape of $H(z)$ entirely driven by the cosmological model.
As a direct consequence, we argue that in spectral siren analyses, it is not trivial to disentangle how much of the observed cosmological constraints truly originate from the GW data itself, as opposed to being a posterior projection induced by the chosen model, in this case, flat-$\Lambda$CDM. 
\begin{figure}[ht]
    \centering
    \includegraphics[width=1\linewidth]{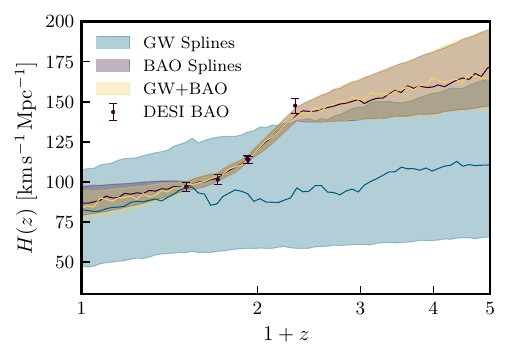}
    \caption{Reconstruction of the inferred Hubble parameter $H(z)$ using our non-parametric model with splines, for GW detections only (blue), BAOs only (purple), and combining GW and BAO (yellow). The BAO data points estimated by DESI are also shown in dark purple on this figure. For each posterior predictive check shown here, the MAP is represented by the solid colored line and the 68.3\% HDI are shown with the contours.}
    \label{fig:Hz_GW_BAO}
\end{figure}

\subsection{Anchoring \texorpdfstring{$H(z)$}{H(z)} with BAOs \label{sec:GW and BAO}}
To further quantify the induced effects of modeling and posterior projections on the estimation of the Hubble parameter $H(z)$, we extend our analysis by incorporating additional ``\textit{anchors}'' that can measure $H(z)$. We include the latest $H(z)$ estimates from BAOs produced by the DESI collaboration, providing five anchor points in the redshift range $0.5 < z < 1.5$. We use the measures of the drag distance $D_\mathrm{M}(z)/r_\mathrm{d}$ from DESI BAO latest measurements and use the sound horizon measured from the CMB to estimate the Hubble parameter \cite{Planck:2018vyg}.
Hence, we now look at three distinct analyses, all using the non-parametric modeling for $H(z)$: the first relies solely on GW data (identical to the left panel of Fig.~\ref{fig:Hz_GW_and_flat_LCDM}), the second uses only the five BAO measurements between redshifts 0.5 and 1.5, and the third combines both GW data and BAO anchors within the same inference framework. The results of these inferences are reported in Fig.~\ref{fig:Hz_GW_BAO}, where the GW measurement is shown in blue, the BAO in purple and the anchored GW with BAO result in yellow. Once again, the MAP is given by the solid line and the colored contours outline the 68.3\% HDIs.

As shown in Fig.~\ref{fig:Hz_GW_BAO}, the five BAO data points from DESI provide robust measurements of the Hubble parameter in the redshift range $0.5 < z < 1.5$. Outside this range, the BAO-only reconstruction of $H(z)$ is clearly model-induced, essentially a posterior projection of the priors inherent to our autoregressive model with splines.

By anchoring our GW-based $H(z)$ reconstruction with BAO measurements, we first observe no additional constraining power in the redshift range $0.5 < z < 1.5$, as expected given the larger error budget of the GW inference in this region.
Below $z < 0.5$, however, we detect a shift in the Hubble parameter toward lower values, accompanied by a slight reduction in overall uncertainties. This effect arises because this redshift range contains the majority of our GW candidates, while the BAO-only inference here is mostly model-induced, as a projection of the prior. Conversely, above $z = 1.5$, the combined anchored reconstruction of $H(z)$ aligns again with the BAO-only results indicating that GW data is not informative at $z>1.5$.

\subsection{Quantifying the GW constraining power \label{sec: constraint power}}

So far, we have demonstrated two key points: \textit{(i)} While it is possible to extract cosmological information from GW data alone, beyond a certain redshift, the reconstruction of the Hubble parameter reduces to a prior-induced projection of the population posterior under the assumed cosmological model, whether a parametric or non-parametric model is used. \textit{(ii)} Anchoring $H(z)$ with external probes such as BAO helps constrain the overall shape of the Hubble parameter and identifies the redshift regions where GWs improve the inferred values, primarily at low redshift where most of the GW candidates lie.

Based on the results presented above, one might consider using the non-parametric inference of $H(z)$ at each node to identify the redshift nodes $z_k$ where GWs provide the tightest constraints on $H(z=z_k)$. However, as we have discussed in Sec.~\ref{sec:GW and BAO} this can not be easily done by using posterior and prior predictive distributions on $H(z)$, as these are implied by the model. To address this, we now turn to the \textit{constraining fraction coefficient} introduced in Eq.~\ref{eq:coef_constraints}, aiming to pinpoint the redshift bin(s) where GW events are most informative for the Hubble parameter. Here, we emphasize that the optimal redshift bin is not the one providing the tightest constraints on $H(z)$, but the one where the estimate of $H(z)$ is least induced by the model and more data-driven by GWs. 
\begin{figure}[ht!]
    \centering
    \includegraphics[width=1.0\linewidth]{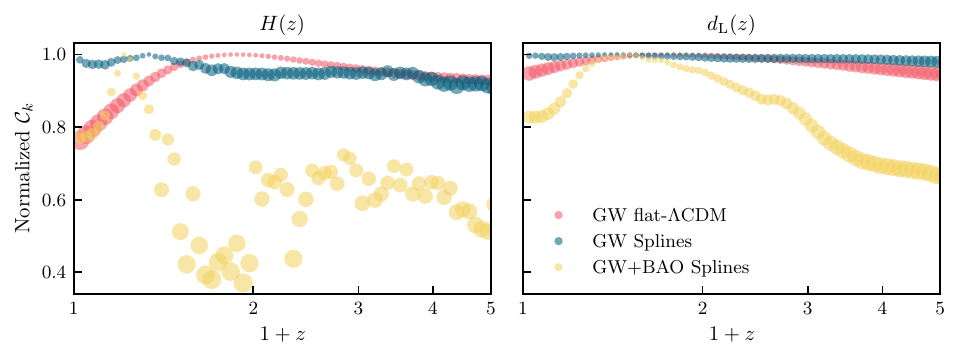}
    \caption{Evolution of the constraining-power coefficient $\mathcal{C}_{k}$ at each redshift nodes $z_{k}$, normalized with respect to their maximum. The lower the size marker, the more data-driven is the constraint at a given node. The \textbf{left} panel shows the constraining power on the Hubble parameter $H(z)$, while the \textbf{right} panel shows the constraining power on the luminosity distance $d_{\rm L}(z)$. Results are obtained with the non-parametric cosmological model with splines, using GWs only (blue), the combination of GWs with BAO (yellow); and with the parametric standard flat-$\Lambda$CDM cosmological model (pink).}
    \label{fig:constraining_power_Hz}
\end{figure}

Figure.~\ref{fig:constraining_power_Hz} shows the estimated constraining fraction coefficients for both the Hubble parameter $H(z)$ (left panel) and the luminosity distance $d_{\rm L}(z)$ (right panel), computed via three approaches: GW-only and GW+BAO with the non-parametric cosmological model; and the strongly parametrized flat-$\Lambda$CDM cosmological model. On the left panel, the constraining coefficient differs significantly between the flat-$\Lambda$CDM and non-parametric models, particularly in the lower redshift bins. This discrepancy arises because in reality, GW candidates are detected in luminosity distance space, not redshift space where $H(z)$ is defined. Consequently, the constraining power on $H(z)$ also reflects how each model, parametric or non-parametric, and their priors, are mapped from the luminosity distance space. The flat-$\Lambda$CDM and non-parametric models thus induce distinct priors on luminosity distance, leading to differing constraints on $H(z)$ across redshift bins. 

We consequently examine the constraining fraction coefficients in a more natural space, the luminosity distance space, $d_{\rm L}(z)$, as shown in the right panel of Fig.~\ref{fig:constraining_power_Hz}. Here, the strong discrepancy in constraining power from GW signals between the non-parametric and parametric approaches using only GW data disappears. Indeed we can see that both the parametrized and non-parametric models provide the same level of constraint from data as function redshift. For the non-parametrized model, the constraining fraction is maximized at at $z=0.44$, although we note that data almost equally contribute to all the redshift bins. The fact that there is no clear data-driven constraint on any of the $z_k$ nodes is a consequence of the fact that GW luminosity distance errors are large, and if coupled with large priors on the $H(z)$ models, they can effectively contribute to any $z_k$ node. 
Finally, we observe that when we include BAO's, the constraining power is clearly maximized in the redshift range $0.5<z<1.5$. This is expected as BAO's provide a direct estimate of $H(z)$ in that range, that calibrates the luminosity distance for GW sources. Consistently with Fig.~\ref{fig:constraining_power_Hz}, we also note that the constraining power is higher for $z<0.5$ and lower for $z>1.5$, indicating that GW sources contributes more to lower redshifts; result in accordance with the typical redshift of GW sources detected to this day.
\begin{figure}[ht!]
    \centering
    \includegraphics[width=1\linewidth]{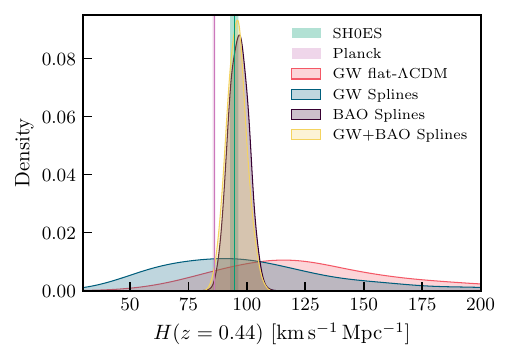}
    \caption{Posterior distributions of the Hubble parameter estimated at $z=0.44$, being the redshift bin with the strongest constraints coming from the GW data. We show results for the three variants of our non-parametric approach with splines: \textsc{GW}, \textsc{BAO}, and \textsc{GW+BAO}. In red, the standard spectral siren approach is used, assuming a flat-$\Lambda$CDM cosmological model. The reference values of $H(z=0.44)$ from SH0ES and Planck are indicated in pink and green, respectively.}
    \label{fig:best_H0}
\end{figure}

Figure.~\ref{fig:best_H0} shows the marginalized posteriors of the Hubble parameter $H(z)$ at $z=0.44$, the redshift bin where GW data provide the strongest constraints. The figure compares results from our non-parametric cosmological model, the flat-$\Lambda$CDM model, and reference values from the SH0ES and Planck collaborations. Using GW data alone with the non-parametric approach, we estimate $H(z=0.44) = 92.3_{-36.6}^{+29.9} \,\hu$, while the flat-$\Lambda$CDM model yields to $H(z=0.44) = 116.3_{-36.6}^{+39.4} \,\hu$ (reporting once again the MAP and 68\% HDI). Even if both values appear to be in good agreement with each other and with other cosmological probes, it is once again evident that the shift between the spline-based reconstruction and the parametric model is created by the choice of model. 
%
We finally note that the constraints on the Hubble parameter are improved for this specific redshift bin, when including BAOs. The non-parametric inference with BAO-only leads to $H(z=0.44) = 97.2_{-3.6}^{+4.8} \,\hu$, while combining with GW data improve it to $H(z=0.44) = 96.3_{-5.1}^{+3.1} \,\hu$.

\section{Conclusions \label{section: conclusion}}
In this article, we present the first non-parametric inference of $H(z)$ using GWs as spectral sirens, representing an important step toward model-independent reconstructions of the Universe’s expansion history. Our analysis shows that inferences of the expansion history from GWs using spectral-siren techniques, even within non-parametric frameworks, remain highly model-dependent, and even more with more traditional parametric models such as $\Lambda$CDM. This highlights the need for caution when interpreting cosmological constraints derived from GW data, including estimates of the Hubble constant, $H_0$. In particular, we identify two issues with the estimation of the Hubble parameter at $z=0$: first, the measurement at this redshift is not the most GW-data-driven, and second, it is influenced by the choice of cosmological model. 

To mitigate model-induced biases, we demonstrate the value of anchoring GW-based measurements with external probes such as with BAOs, which help identify the redshift regions where GWs provide meaningful constraints. Furthermore, we show how the constraining power can be quantified directly in luminosity-distance space, introducing a novel estimator that identifies the redshift bins most strongly constrained by GW data. Our results indicate that, as of today and using the GWTC-4.0 catalog for spectral siren, the redshift bin centered at $z = 0.44$ yields the most robust estimate of $H(z=0.44) = 92.3_{-36.6}^{+29.9} \,\hu$.

Finally, we emphasize a crucial consideration for the field of GW cosmology and future works: all GW analyses employing Bayesian inference for cosmology purposes, particularly spectral siren studies, must explicitly acknowledge and ideally discuss the influence of model-induced constraints, especially when parametric representations are employed to describe the cosmological model.

\section{Acknowledgments}
This work is supported by ERC grant GravitySirens  101163912. Funded by the European Union. Views and opinions expressed are however those of the author(s) only and do not necessarily reflect those of the European Union or the European Research Council Executive Agency. Neither the European Union nor the granting authority can be held responsible for them. This material is based upon work supported by NSF's LIGO Laboratory which is a major facility fully funded by the National Science Foundation.

\section*{Appendix \label{app:pop models}}
\subsection*{Details on the mass distribution of the binary black holes}
In this paper, to describe the the distribution of binary black hole masses, we selected on parametric model: the \textsc{Broken Power Law + 2 Peaks}. As explained in section.~\ref{section: framework}, this model inspired from \cite{LIGOScientific:2025pvj} is constructed from the combination of a broken power law $B(m_{1,\rm s}|\boldsymbol{\Lambda})$ breaking at $m_{\rm break}$ and two Gaussian components $G_{1}(m_{1,\rm s}|\boldsymbol{\Lambda})$ and $G_{2}(m_{1,\rm s}|\boldsymbol{\Lambda})$.

The probability density function of the broken power law is given by 
\begin{equation}
    B(m_{1,\rm s}|\boldsymbol{\Lambda})= \frac{1}{N}
    \begin{cases}
     \big(\frac{m_{1,\rm s}}{m_{\rm break}} \big)^{-\alpha_1} &  m_{\rm min}\leq m_{1,\rm s} < m_{\rm break} \\
    \big(\frac{m_{1,\rm s}}{m_{\rm break}} \big)^{-\alpha_2} &  m_{\rm break}\leq m_{1,\rm s} < m_{\rm max}\,,
    \end{cases}
\end{equation}
where N is a normalization constant, and $\alpha_1$ and $\alpha_2$ are the index of each power laws before and after the break.

The Gaussian peaks are defined such that
\begin{eqnarray}
    G_i(m_{1,\rm s}|\boldsymbol{\Lambda}) = \mathcal{N}(m_{1,\rm s}|\mu_{i},\sigma_{i},\rm low, high),
\end{eqnarray}
where $\mathcal{N}$ is a truncated Gaussian between a low and high value.

\subsection*{Details on the hyper-parameters and prior ranges}
\begin{table}[ht!]
\caption{Summary of the cosmological and population parameters, along their hyper-priors used for the Bayesian inference. The first column shows the parameter symbol, the second column its description in our models, and the third column the prior range.}
\centering
\begin{tabular}{l c c}
\\
\multicolumn{3}{c}{\textbf{$\boldsymbol{H(z)}$ autoregressive spline model}} \\
\hline
Population parameter & Description & Prior \\
\hline
$\ln H_0$ & Logarithm of the Hubble parameter at $z=0$ in [\hu] & $\mathcal{U}[\ln 20,\ln 250]$  \\
$w_i$  & EOS coefficients in the i-th redshift bin & $\mathcal{U} [-2,1/3]$   \\ \\
\hline
\hline \\

\multicolumn{3}{c}{\textbf{$\boldsymbol{H(z)}$ flat-$\boldsymbol{\Lambda}$CDM model}} \\
\hline
Population parameter & Description & Prior \\
\hline
$H_0$ & Hubble constant in [\hu] & $\mathcal{U}[20,250]$  \\
$\Omega_{\rm m,0}$  & Fractional energy density in matter at $z=0$ & $\mathcal{U}[0,1]$ \\ \\
\hline
\hline \\

\multicolumn{3}{c}{\textbf{Broken Power Law + 2 Peaks mass model}} \\
\hline
Population parameter & Description & Prior \\
\hline
$m_{\rm break}$ & Transition point in [\Msol] between the two power laws& $\mathcal{U}[20,50]$  \\
$\alpha_1$ & low-mass power law slope & $\mathcal{U}[-3,2]$  \\
$\alpha_2$ & High-mass power law slope & $\mathcal{U}[-6,2]$  \\
$m_{\rm max}$ & Maximum BBH mass in [\Msol] & $\mathcal{U}[80,200]$  \\
$\delta_{m,\rm low}$ & low-mass tapering window in [\Msol] & $\mathcal{U}[80,200]$  \\
$\delta_{m,\rm high}$ & low-mass tapering window in [\Msol] & $\mathcal{U}[80,200]$  \\
$\mu_{g,\rm low}$ & Mass of the lower gaussian peak in [\Msol] & $\mathcal{U}[5,20]$  \\
$\mu_{g,\rm high}$ & Mass of the lower gaussian peak in [\Msol] & $\mathcal{U}[25,60]$  \\
$\sigma_{g,\rm low}$ & Width of the lower gaussian peak in [\Msol] & $\mathcal{U}[0.1,10]$  \\
$\sigma_{g,\rm high}$ & Width of the lower gaussian peak in [\Msol] & $\mathcal{U}[0.1,10]$  \\
$\lambda_{\rm g}$ & Fraction of events in the gaussian peaks & $\mathcal{U}[0.1,10]$  \\
$\lambda_{\rm g, low}$ & Fraction of events in the lower gaussian peak & $\mathcal{U}[0.1,10]$  \\ \\
\hline
\hline \\

\multicolumn{3}{c}{\textbf{Mass ratio model}} \\
\hline
Population parameter & Description & Prior \\
\hline
$\beta$ & Power law slope & $\mathcal{U}[-1,5]$  \\
$\Omega_{\rm m,0}$  & Fractional energy density in matter at $z=0$ & $\mathcal{U}[0,1]$\\ \\
\hline
\hline \\

\multicolumn{3}{c}{\textbf{Merger rate model}} \\
\hline
Population parameter & Description & Prior \\
\hline
$z_p$ & Peak of BBH merger rate in redshift & $\mathcal{U}[0,5]$  \\
$\gamma$ & Power low slope for $z<z_p$ & $\mathcal{U}[0,5]$  \\
$k$ & Power low slope for $z>z_p$ & $\mathcal{U}[0,5]$ \\ \\
\hline
\hline \\

\end{tabular}
\label{tab:1}
\end{table}

\bibliographystyle{ieeetr}
\bibliography{bright_sirens,analysis_techniques,dark_sirens,detector_references,EM_surveys,modified_gravity_theory,population_properties,cosmology_references,gwtc4_lvk_ref}

\end{document}